\documentclass[11pt]{llncs}
\usepackage{fullpage}
\usepackage{makeidx}  
\usepackage{alltt}
\usepackage[english]{babel}
\usepackage{verbatim}
\usepackage{t1enc}
\usepackage{float}
\usepackage{ifpdf}
\usepackage{amsmath}
\usepackage{amssymb}
\usepackage{qtree}
\usepackage{appendix}
\usepackage{listings}
\usepackage{framed}
\usepackage{caption}

\ifpdf
  \usepackage[pdftex]{graphicx}
\else
  \usepackage[dvips]{graphicx}
\fi

\usepackage[bookmarks=true,
            a4paper=true,
            colorlinks=true,
            linkcolor=blue,
            bookmarksnumbered=true,
            pdfpagemode=UseOutlines,
            pagebackref=true]
            {hyperref}

\usepackage{algcompatible}
\usepackage{algorithm}
\usepackage{caption}
\begin{document}
%
%
\pagestyle{headings}  

%
%
\title{A space-- and time--efficient Implementation of the Merkle Tree Traversal Algorithm}
\titlerunning{Merkle's tree traversal}  
%
\author{Markus Knecht \and Willi Meier
 \and Carlo U. Nicola}

\authorrunning{Markus Knecht et al.} 
%
\tocauthor{Markus Knecht, Willi Meier and Carlo U. Nicola}
\institute{University of Applied Sciences Northwestern Switzerland, School of Engineering, \\ 5210 Windisch , Switzerland. \\
\email{\{markus.knecht,willi.meier,carlo.nicola\}@fhnw.ch}}

\maketitle              

\begin{abstract}
We present an algorithm for the Merkle tree traversal problem which combines the efficient space-time trade-off from the fractal Merkle
tree~\cite{key3} and the space efficiency from the improved log space-time Merkle trees traversal~\cite{key15}. We give an exhaustive analysis of the
space and time efficiency of our algorithm in function of the parameters $H$ (the height of the Merkle tree) and $h$ ($h = \frac{H}{L}$ where $L$ is
the number of levels in the Merkle tree). We also analyze the space impact when a continuous deterministic pseudo--random number generator (PRNG) is
used to generate the leaves. We further program a low storage--space and a low time--overhead version of the algorithm in Java and measure its
performance with respect to the two different implementations cited above.  Our implementation uses the least space when a continuous PRNG is used
for the leaf calculation.

\end{abstract}

\section{Introduction}
Merkle's binary hash trees are currently very popular, because their security is independent from any number theoretic conjectures~\cite{key13}.
Indeed their security is based solely on two well defined properties of hash functions: (i) Pre-image resistance: that is, given a hash value $v$, it
is difficult to find a message $m$ such that $v=hash(m)$. The generic pre-image attack requires $2^n$ calls to the hash function, where $n$ is the
size of the output in bits. (ii) Collision resistance: that is, finding two messages $m_1 \neq m_2$ such that $hash(m_1)=hash(m_2)$ is difficult. The
generic complexity of such an attack is given by the birthday bound which is $2^{n/2}$ calls to the hash function. It is interesting to note that the
best quantum algorithm to date for searching $N$ random records in a data base (an analogous problem to hash pre-image resistance) achieves only a
speedup of $\mathcal{O}(\sqrt{N})$ to the classical one $\mathcal{O}(N)$~\cite{key1}. More to the point in~\cite{key1.1} the speedup of a quantum
algorithm that finds collisions in arbitrary r-to-one functions is $\mathcal{O}(\sqrt[3]{N/r})$ to the classical one.

\noindent A Merkle tree is a complete binary tree with a $n$-bit hash value associated with each node. Each internal node value is the result of a
hash of the node values of its children. Merkle trees are designed so that a leaf value can be verified with respect to a publicly known root value
given the authentication path of the respective leaf. The authentication path for a leaf consists of one node value at each level $l$, where $l = 0,
\cdots, H-1$, and $H$ is the height of the Merkle tree ($H \leq 20$ in most practical cases). The chosen nodes are siblings of the nodes on the path
connecting the leaf to the root.

\noindent The Merkle tree traversal problem answers the question of how to calculate efficiently\footnote{The authors of~\cite{key14} proved that the
bounds of space ($\mathcal{O}(t H/log(t))$) and time ($\mathcal{O}(H/log(t))$)) for the output of the authentication path of the current leaf are
optimal (t is a freely choosable parameter).} the authentication path for all leaves one after another starting with the first $Lea\!f_0$ up to the
last $Lea\!f_{2^H-1}$, if there is only a limited amount of storage available (e.g. in Smartcards ).

\noindent The generation of  the root of the Merkle tree (the public key in a Merkle signature system) requires the computation of all nodes in the
tree. This means a grand total of $2^H$ leaves' evaluations and of $2^{H} - 1$ hash computations. The root value (the actual public key) is then
stored into a trusted database accessible to the verifier.

\noindent The leaves of a Merkle tree are used either as a one--time token to access resources or as building block for a digital signature scheme.
Each leaf has an associated private key that is used to generate either the token or a signature building block (see \ref{sec:prng}). The tokens can
be as simple as a hash of the private key. In the signature case, more complex schemes are used in the literature (see for example~\cite{key2.1} for
a review).

\subsection*{Related work}
Two solutions to the Merkle tree traversal problem exist. The first is built on the classical tree traversal algorithm but with many small
improvements~\cite{key15} (called log algorithm from now on). The second one is the fractal traversal algorithm~\cite{key3} (called fractal algorithm
from now on). The fractal algorithm trades efficiently space against time by adapting the parameter $h$ (the height of both  $Desired$ and $Exist$
subtrees, see Fig.~\ref{fig:2}), however the minimal space it uses for any given $H$ (if $h$ is chosen for space optimality) is more than what the
log algorithm needs. The log algorithm cannot as effectively trade space for performance. However, for small $H$ it can still achieve a better time
and space trade-off than the fractal algorithm.

\noindent A study~\cite{key14} analyses theoretically the impact on space and time--bounds of some enhancements to both the log and fractal
algorithm, which are important to our implementation.

\subsection*{Our contributions}
We developed an algorithm for the Merkle tree traversal problem which combines the efficient space-time trade-off from~\cite{key3}  with the space
efficiency from~\cite{key15}. This was done by applying all the improvements discussed in~\cite{key15} to the fractal algorithm~\cite{key3}. We have
also analyzed the space impact of a continuous deterministic pseudo--random number generator (PRNG)\footnote{A deterministic pseudo--random number
generator which can not access any random number in its range without first computing all the preceding ones.} on the algorithms. All these
improvements lead to an algorithm with a worst case storage of $[L \times 2^h +2H-2h]$ hash values (Sec.~\ref{sec:st_summ}). The worst case time
bound for the leaves' computation per authentication path, amounts to $\frac{2^h -1}{2^h} \times (L-1) + 1$ (Sec.~\ref{sec:st_summ}). This means a
reduction in space of about a factor 2 compared with the fractal algorithm~\cite{key3} (see Fig.~\ref{fig:5} and Fig.~\ref{fig:6}).

\noindent Although on first sight our enhancements are predated by~\cite{key14} three main differences distinguish our contribution
vis-à-vis~\cite{key14}: (i) Our use of a different $TreeHash$ and metrics; (ii) Our special computation of the $Desired$ tree (see
Section~\ref{sec:des_an}) and (iii) Our use of a continuous PRNG in leaf computation.

\noindent We further implemented the algorithm in Java with focus on a low space and time overhead~\cite{key0} and we measured its performance
(Sec.~\ref{sec:res}).

\section{Preliminaries}
\begin{figure}[htb]
    \centering
    \includegraphics*[scale=1.5, keepaspectratio=true]{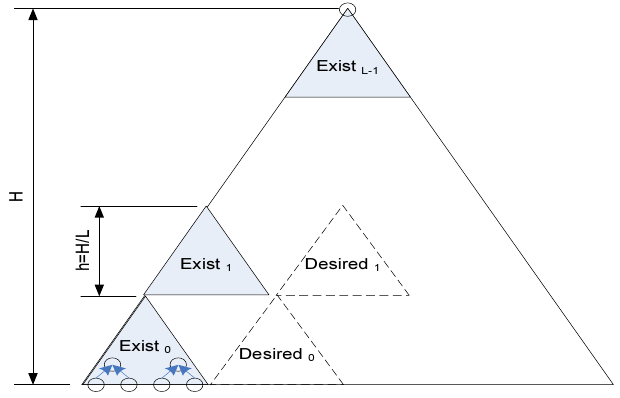}
    \caption{Fractal Merkle tree structure and notation (Figure courtesy of~\cite{key3}). A hash tree of height $H$ is divided into $L$ levels,
    each of height $h$. The leaves of the hash tree are indexed $\{0, 1, ..., 2^H - 1\}$ from left to right. The height of a node is defined as
    the height of the maximal subtree for which it is the root and ranges from $0$ (for the leaves) to $H$ (for the root).
    An $h$-subtree is
    "at level $i$" when the height of its root is $(i+1)h$ for some $i \in \{0, 1,\cdots,L-1\}$.}
    \protect\label{fig:2}
\end{figure}

The idea of the fractal algorithm~\cite{key3} is to store only a limited set of subtrees within the whole Merkle tree (see Fig.~\ref{fig:2}). They
form a stacked series of $L$ subtrees $\{Subtree_i\}_{i=0}^{L-1}$. Each subtree consists of an $Exist$ tree $\{Exist_i\}$ and a $Desired$ tree
$\{Desired_i\}$, except for $Subtree_{L-1}$, which has no $Desired$ tree. The $Exist$ trees contain the authentication path for the current leaf.
When the authentication path for the next leaf  is no longer contained in some $Exist$ trees, these are replaced by the $Desired$ tree of the same
subtree. The $Desired$ trees are built incrementally after each output of the authentication path algorithm, thus minimizing the operations needed to
evaluate the subtree.

\noindent The nodes in a Merkle tree are calculated with an algorithm called $TreeHash$.  The algorithm takes as input a stack of nodes, a leaf
calculation function and a hash function and it outputs an updated stack, whose top node is the newly calculated node. Each node on the stack has a
height $i$ that defines on what level of the Merkle tree this node lies: $i = 0$ for the leaves and $i = H$ for the root. The $TreeHash$ algorithm
works in small steps. On each step the algorithm looks at its stack and if the top two elements have the same height it pops them and puts the hash
value of their concatenation back onto the top of stack which now represents the parent node of the two popped ones. Its height is one level higher
than the height of its children. If the top two nodes do not have the same height, the algorithm calculates the next leaf and puts it onto the stack,
this node has a height of zero.

\begin{figure}[htb]
    \centering
    \includegraphics[scale=0.3, keepaspectratio=true]{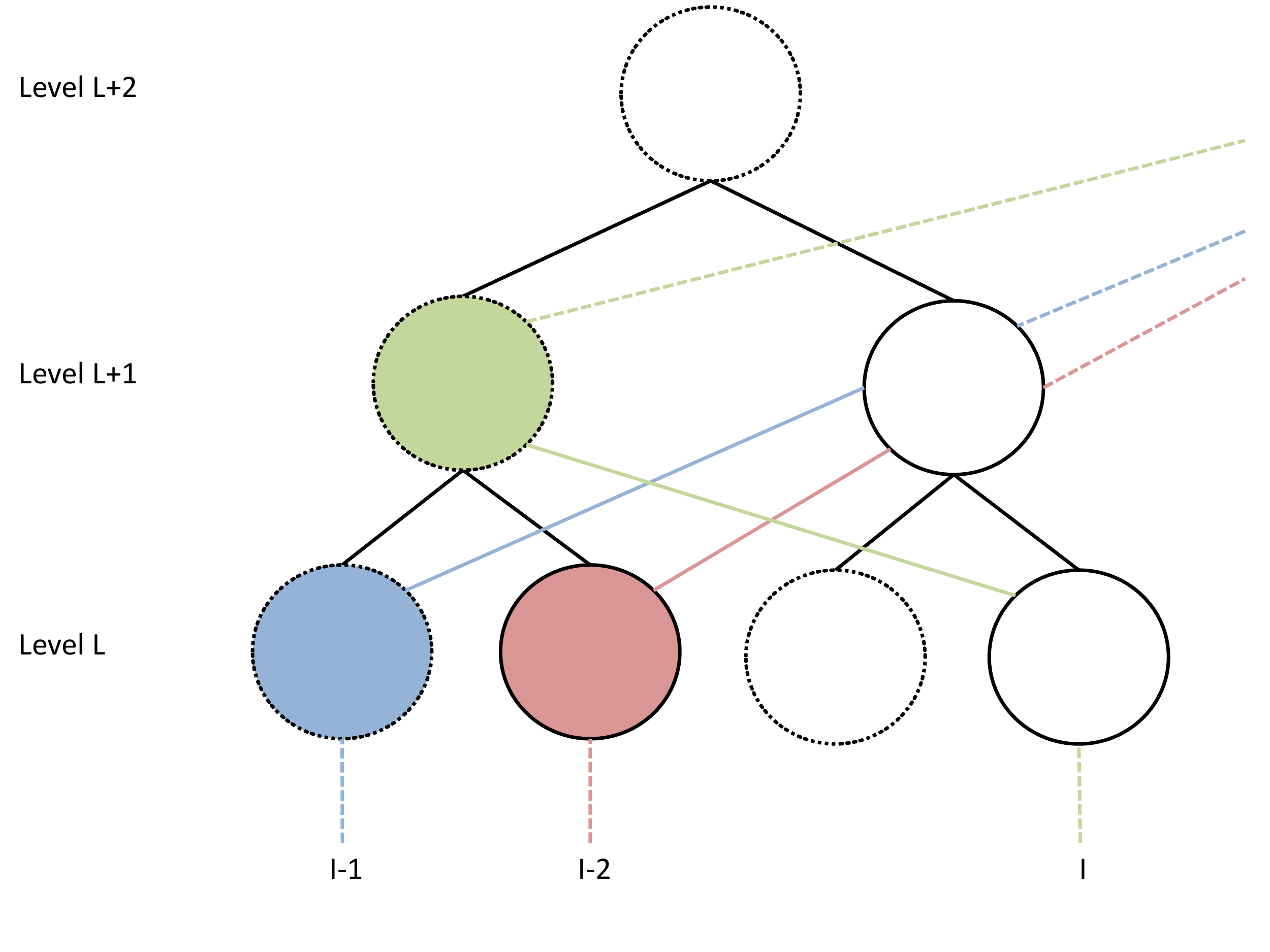}
    \caption[Authentication path changes on a level $L$]{The colored lines mark the different authentication paths. The index $I$ at the start of each line indicates how many times a node on level $L$
     of the authentication path has changed. An authentication path whose node has changed $I$ times on level $L$ has changed $I \times (2^L)$ times on level $0$ (which changes each round). The
    dotted circles are left nodes or the root of a subtree which are not stored in a subtree.}\
    \protect\label{fig:8}
\end{figure}

We quickly summarize the three main areas where our improvements were critical for a better space--time performance of the original fractal
algorithm:
\begin{enumerate}

    \item Left nodes have the nice property, that when they first appear in an authentication path,
    their children were already on an earlier authentication path (see Fig.~\ref{fig:8}).
    For right nodes this property does not hold. We can use this fact to calculate left
    nodes just before they are needed for the authentication path without the need to store them in the subtrees.
    So we can save  half of the space needed for the subtrees, but compared to the fractal algorithm one additional leaf calculation has to be carried out every two rounds (one round corresponds to the calculation of one authentification path).

    \item In most practical applications, the calculation of a leaf is more expensive than the calculation of an inner node\footnote{A inner node is a node with height greater than zero.}. This can be used to design
    a variant of the $TreeHash$ algorithm, which has a worst case time performance that is nearer to its
    average case for most practical applications. The improved $TreeHash$ (see Algorithm~\ref{algo:treehash}) given one leaf, calculates as many inner nodes
    as possible per update (see Section~\ref{sec:metr}) before needing a new leaf, instead of processing just one leaf or one inner node as in the normal case.

    \item In the fractal Merkle tree one $TreeHash$ instance per subtree exists for calculating the nodes of the $Desired$ trees and each of
    them gets two updates per round. Therefore all of them have nodes on their stacks which need space of the order of $\mathcal{O}(\frac{H^2}{h})$.
        We can distribute the updates in another way, so that the associated stacks are mostly empty~\cite{key8}.
    This reduces the space needed by the stacks of the $TreeHash$ instances to $\mathcal{O}(H-h)$.
\end{enumerate}

It is easy enough to adapt point one and two for the fractal algorithm, but point three needs some changes in the way the nodes in a
subtree are calculated (see Sec.~\ref{sec:des_an}).
\begin{algorithm}[h]
  \caption{Listing: Generic version of $TreeHash$ that accepts different types of $\mathrm{Process_i}$ (See Appendix~\ref{app} for a thorough definition of $\mathrm{Process_i}$).
    A node has a height and an index, where the index indicates where a node is positioned in relation to all nodes with the same height in the Merkle tree.
    }
  \begin{framed}
        \begin{algorithmic}

        \STATE $ \mathbf{INPUT}: \mathrm{StackOfNodes}, \mathrm{Leaf}, \mathrm{Process_i}, \mathrm{SubtreeIndex}$
        \STATE $ \mathbf{OUTPUT}: \mathrm{updated\:StackOfNodes}$
        \STATE $\mathrm{Node} \leftarrow  \mathrm{Leaf}$
        \IF{$\mathrm{Node.index} \pmod 2 == 1$} \STATE $continue \leftarrow \mathrm{Process_i(Node,SubtreeIndex)}$
        \ELSE
            \STATE $continue \leftarrow 1$
        \ENDIF
        \WHILE{$continue \neq 0 \wedge (\mathrm{Node.height} == \mathrm{StackOfNodes.top.height})$}
            \STATE{$\mathrm{Node} \leftarrow \mathrm{hash(StackOfNodes.pop || Node)}$}
            \STATE $continue \leftarrow \mathrm{Process_i(Node,SubtreeIndex)}$
        \ENDWHILE
        \IF{$continue \neq 0$}
            \STATE $\mathrm{StackOfNodes.push(Node)}$
        \ENDIF

\end{algorithmic}\label{treehash}
\end{framed}
\label{algo:treehash}
\end{algorithm}

\section{Algorithm's overview}\label{sec:overview}
In this Section we will give an overview of the complete algorithm and explain how all its components work together. The algorithm is divided into
two phases. The first phase is the initialisation phase in which the public key is calculated (see Alg.~\ref{algo:init}). We run in this phase the
improved $TreeHash$(see Alg.~\ref{algo:treehash}) from~\cite{key15}, step by step until the root node is computed. The improved $TreeHash$ algorithm
needs $Lea\!f_i$ where $i \in \{0, 1,\cdots,2^H-1\}$ as an input.

\noindent The value of $Lea\!f_i$ is dependent on the usage of the Merkle tree. It could be as simple as a token, where the leaf is a hash of the
tokens private key, or a one time signature scheme like Winternitz~\cite{key13} where the leaf is the public key of the one--time signature. The
private keys needed to compute the leaves are provided by a PRNG, whose key  corresponds to the private key of the complete Merkle tree.

In the initialisation phase each node is computed exactly once. This fact is used to store all right nodes in the first $Exist$ tree of each subtree
and all the nodes in the authentication path for the $Lea\!f_0$. The second phase iteratively generates the authentication paths for all the
remaining $Lea\!f_i$ (from left to right) where $i \in \{1, 2,\cdots,2^H-1\}$ (see Alg.~\ref{algo:1} and Alg.~\ref{algo:2}). Each authentication path
can be computed by changing the previous one~\cite{key13}. The authentication path for $Lea\!f_i$ changes on a level $k$ if $2^k|i$. If the node
changes to a right node, it can be found in one of the $Exist$ trees. If it changes to a left node, it can be computed from its two children. The
right child can be found in the $Exist$ trees and the left child is on the previous authentication path (see Fig.~\ref{fig:8}).

When a node in the $Exist$ tree is no longer needed for the computation of any upcoming authentication path, it is removed. To prevent the $Exist$
tree running out of nodes, all the nodes in the $Desired$ tree have to be computed before the $Exist$ tree has no nodes left. This is done with the
help of two $TreeHash$ instances per subtree. One, called the lower $TreeHash$, calculates all nodes on the bottom level\footnote{The bottom level is
the lowest level in a $Desired$ tree.} of a $Desired$ tree (called bottom level nodes) from the leaves of the Merkle tree. The other, called the
higher $TreeHash$, calculates all the remaining $Desired$ nodes\footnote{$Desired$ nodes are all the nodes stored in a $Desired$ tree.} (called
non-bottom level nodes) from the bottom level ones. All the lower $TreeHash$ instances use the same scheduling algorithm as in~\cite{key15} with
$L-1$ updates per round. The higher $TreeHash$ uses a custom scheduling algorithm which executes an update every $2^{bottomLevel}$ rounds. The higher
$TreeHash$ produces a node on a level $k$ in the $Desired$ tree every $2^k$ rounds, which corresponds to the rate at which the authentication path
changes on that level. When the last node from the $Exist$ tree is removed, all the nodes in the $Desired$ tree are computed and the $Exist$ tree can
be replaced with the $Desired$ tree. In section~\ref{sec:node_ready} we will prove that the lower $TreeHash$es produce the nodes on the bottom level
before the higher $TreeHash$es need them, if $L-1$ updates are done per round. A lower $TreeHash$, which has terminated, is initialized again as soon
as the generated node is used as input for the higher $TreeHash$. Because only the right nodes are stored in the subtrees, the $TreeHash$es do only
have to compute right nodes and those left nodes which are used to calculate a right node contained in the $Desired$ tree. The only left nodes never
used to compute a right node in a $Desired$ tree are the first left node at each level in each $Desired$ tree. To ensure that no unneeded nodes are
computed, the lower $TreeHash$ does not compute nodes for its first $2^{bottomLevel}$ updates per $Desired$ tree and so does the higher $TreeHash$
for its first update per $Desired$ tree. These skipped updates are nevertheless counted without the scheduling algorithm assigning them to another
$TreeHash$. Therefore from the point of view of the scheduling algorithm the $TreeHash$ behaves as if the nodes would have been computed.

\section{Analysis}

\subsection{\textsl{TreeHash} Metrics}\label{sec:metr}
Below we give some definitions that will firstly permit a better understanding of our analysis and secondly, unify all the similar  concepts
scattered in the literature. We define as classical \textsl{TreeHash} the algorithm used in~\cite{key3}. In that paper a $step_C$ is defined as  the
calculation of either one leaf or the node's hash. We define as improved \textsl{TreeHash} the algorithm used in~\cite{key15}. Therein a $step_I$ is
defined as the calculation of the sum of one leaf and X hashes (where X is the number of nodes' computations before a new leaf is needed). We define
$update = 2 \times step_C$ in the case of the classical \textsl{TreeHash} and $update = 1 \times step_I$ in the case of the improved
\textsl{TreeHash}.

Since in our work we assume that the hash computation time is very small compared to a leaf's computation (an assumption certainly valid for MSS
(Merkle Signature Scheme)), we use as the basic time unit (metrics) for this work  the time we need to compute a leaf.

So we can claim that in the worst case condition a classical \textsl{TreeHash} update takes 2 leaves' computations whereas an improved
\textsl{TreeHash} update takes only one. For the calculation of all nodes in a tree of height $H$, the classical $TreeHash$ needs $2^H - 2^{-1}$
updates ($2^{-1}$ because the last update needs only to do one $step_C$). On the contrary the improved $TreeHash$ needs $2^H$ updates to reach the
same goal.

\subsection{Computation of the Desired tree}\label{sec:des_an}
In this section we will explain how the nodes in a $Desired$ tree are computed and stored which  hallmarks the main difference of our algorithm to
the algorithm in~\cite{key14}. Recall that in~\cite{key14} the $TreeHash$ algorithm of a $Desired$ tree gets $2^h \times (2^{bottomLevel})$ updates
for calculating its nodes. The $2^h$ bottom level nodes of a $Desired$ tree are calculated during the first $2^h \times (2^{bottomLevel} - 2^{-1})$
updates \footnote{Remember that a classical $TreeHash$ needs $2^{bottomLevel} -2^{-1}$ updates to compute a bottom level node. Furthermore in all our
derivation $h|H$ and $h \leq H$ hold.}. After calculating the bottom level nodes there are $2^{h-1}$ updates left ( from: $(2^h \times
(2^{bottomLevel})) - (2^h \times (2^{bottomLevel} - 2^{-1})) = 2^{h-1}$). These are used to calculate the non-bottom level nodes of the $Desired$
tree~\cite{key14}. There is no additional space needed to calculate the non-bottom level nodes from the bottom level nodes. This because after
calculating a new node, the left child is dropped and the new value can be stored instead~\cite{key14}. This approach can not be used with the
improved $TreeHash$ from~\cite{key15} without increasing the amount of updates the $TreeHash$ of a $Desired$ tree gets before the $Desired$ tree has
to be finished. This is due to the fact, that the improved $TreeHash$ needs $2^h \times (2^{bottomLevel})$ updates to compute all bottom level nodes
of the $Desired$ tree, which would leave 0 updates for the calculation of the non-bottom level nodes.

\noindent As described in Section~\ref{sec:overview} our algorithm uses a lower $TreeHash$ and a higher $TreeHash$ per subtree. All the lower
$TreeHash$ instances use the same scheduling algorithm as in~\cite{key15} with $L-1$ updates per round. The higher $TreeHash$es use a custom
scheduling algorithm which executes an update every $2^{bottomLevel}$ rounds. The main difference vis-à-vis~\cite{key14} is that we compute the nodes
in the $Desired$ tree continuously during the calculations of the $Desired$ tree, and not only at the end. This approach distributes the leaf
computations during the computation of a $Desired$ tree more equally than the one from~\cite{key14}.

\subsubsection{Space analysis for Desired tree computation}\label{sec:pace_des}
We will show that our algorithm needs $L\times (2^h-1)$ hash values for the $Exist$ and $Desired$ tree, when the authentication path is taken into
account, instead of $L\times(2^h)$ hash values needed by the algorithm in~\cite{key14}.

\noindent The authentication path is a data structure which can store one node per level. Because the authentication path is contained in all the
$Exist$ trees (which store only right nodes), right nodes on the authentication path are contained in both structures and thus have to be held only
once in  memory.

\noindent The authentication path changes on a level $k$ every $2^{k}$ rounds and the higher $TreeHash$ produces a node on a level $k'$ every
$2^{k'}$ rounds. Whenever a left node enters the authentication path, its right sibling leaves the authentication path and can be discarded (with one
exception discussed below). From this we can conclude (ignoring the exception for now), that every $2^{k+1}$ rounds the $Exist$ tree discards a right
node on level $k$ and the higher $TreeHash$ produces a left node on the same level. This means the higher $TreeHash$ can store one left node on each
level using the space of the discarded nodes in the $Exist$ tree. The right nodes the higher $TreeHash$ produces can be stored using the space of the
left node from which they have been computed.

\noindent We will now look at the exception mentioned above: a right node on level $k$ which has a left node as parent, cannot be discarded when it
leaves the authentication path, because it is needed for the computation of its parent as explained in~\cite{key15}. It will be discarded $2^k$
rounds after it left the authentication path. During these $2^k$ rounds there can be a left node with height $k$ on the higher $TreeHash$, for which
fresh storage space must be provided. Fortunately this situation can only occur if there is a right node on the authentication path (the sibling of
the parent of the node which could not be discarded). This right node is stored in both the $Exist$ tree and the authentication path and must be held
in memory only once.

\noindent The special scheduling of the lower $TreeHash$ (see Sec.~\ref{sec:tree_hash}) may compute a node on the bottom level that is not
immediately consumed by the higher $TreeHash$ and therefore should be stored until needed. We can store this node in the space reserved for the higher
$TreeHash$, because the left node with the highest level on a higher $TreeHash$ is never stored, for the simple reason that it is not needed for
the calculation of any right node in the $Desired$ tree (see Fig.~\ref{fig:3}).

\noindent From this we conclude that the authentication path and all the subtrees together use no more than $L\times (2^h-1)+H$ space, where $h$ is
the height of a subtree, ($2^h-1$) is the amount of nodes a tree of height $h$ needs, when it stores only right nodes (see Fig.~\ref{fig:3}) and $H$
is the space needed to store the current authentication path.

\subsubsection{Sharing the same Data structure in both Exist and Desired Trees}\label{sec:share_ex_des}
We now show that we can store the nodes of the $Exist$ tree and the $Desired$ tree in one single tree data structure. This is the case, because we
can store two related\footnote{Two nodes of either a $Desired$ or an $Exist$ tree are said to be related if they have the same position relative to
their root.} nodes in the same slot. We can do this because when a node in the $Desired$ tree is stored, its related node in the $Exist$ tree was
already discarded. This is trivial for left nodes, because they are never stored in the $Exist$ or $Desired$ tree. In the previous section we showed
that with one exception, the $Exist$ tree discards a right node on a level in the same round the higher $TreeHash$ computes a left node on that
level. The sibling of the left node a higher $TreeHash$ computes every $2^{l+1}$ rounds on a level $l$, is the node related  to the right node the
$Exist$ tree discards during this round. The right node which is computed $2^{l}$ rounds later on the level $l$ is the node related to the discarded
one and so it can be stored in the same slot of the data structure. We now look at the special case: right nodes with a left node as parent (see
Sec.~\ref{sec:pace_des}). Such a right node on level $k$ will be discarded $2^k$ rounds later than the other right nodes. It will be discarded in the
same round as the higher $TreeHash$ produces its related node. We ensure that the slot in the data structure is free by calculating left nodes in the
authentication path before we update the higher $TreeHash$ (see Algorithm~\ref{algo:1}).

\begin{figure}[h]
    \centering
    \includegraphics*[scale=0.4, keepaspectratio=true]{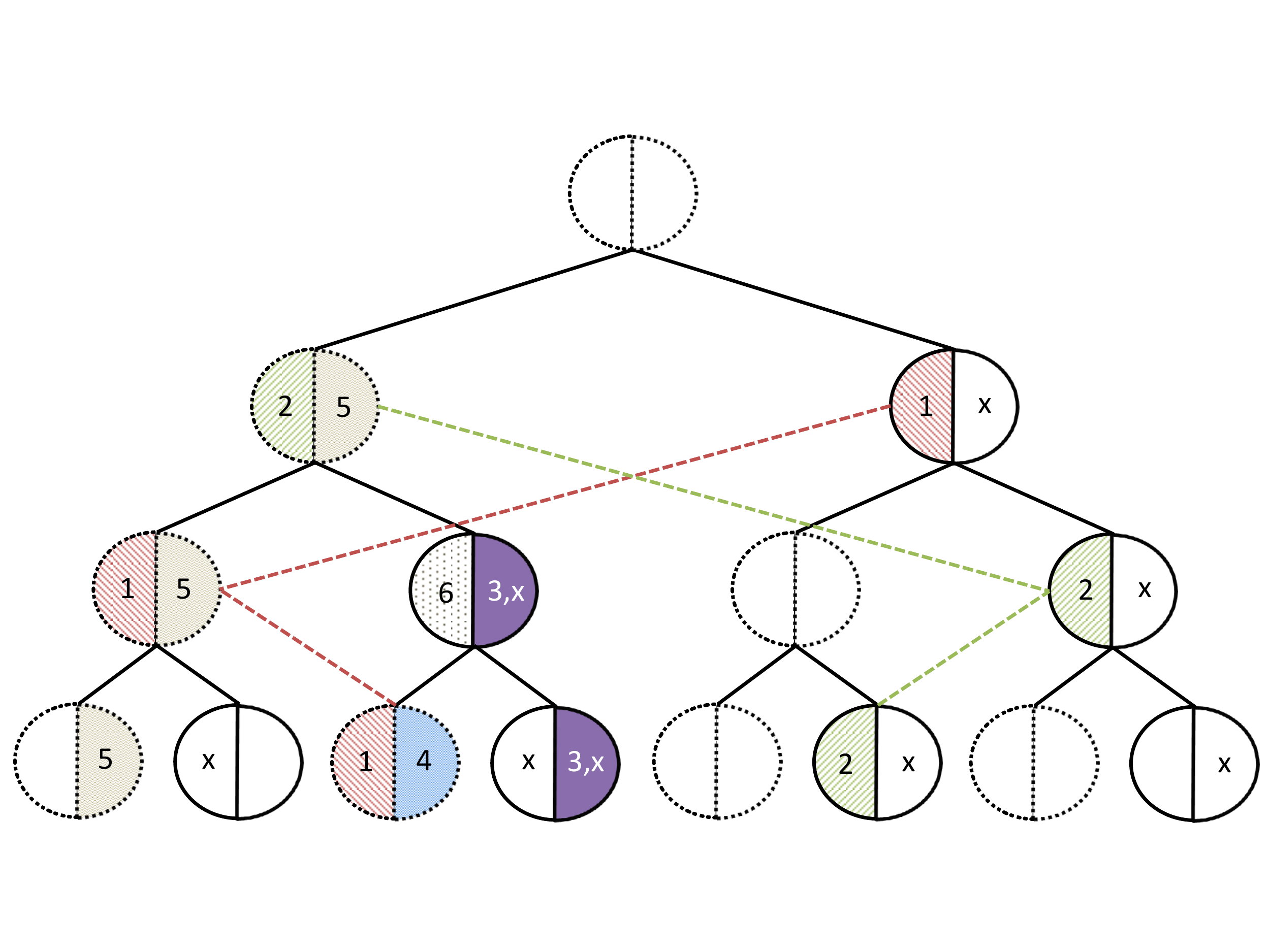}
    \caption{The left half of each circle represents the $Exist$ tree and the right half the $Desired$ tree. The nodes with dotted lines are left nodes or the root and thus are not stored in the subtree,
    but they may be stored in the authentication path or on the higher $TreeHash$. The markings on the nodes have the following meanings:
    Label $x$: nodes already discarded (in case of $Exist$ tree) or not yet computed (in case of $Desired$ tree).
    Label $1$: nodes lying on the current authentication path.
    Label $2$: nodes lying on the upcoming authentication path.
    Label $3$: nodes computed next by the higher $TreeHash$.
    Label $4$: left node on the higher $TreeHash$.
    Label $5$: left nodes which do never contribute to a right node calculation (not stored in higher $TreeHash$).
    Label $6$: node which could not yet be discarded, because it is needed for calculating a left node in the upcoming authentication path.}\
    \protect\label{fig:3}
\end{figure}

In Fig.~\ref{fig:3} we show how the different nodes of the $Desired$ and $Exist$ trees are managed.

\subsubsection{Space used for the key generation of the leaves}~\label{sec:prng}
In this section we will analyse the space used by the deterministic PRNG, which calculates the private keys used in the leaf calculations. Supposing
the PRNG algorithm can generate any random number within its range without first calculating all the preceding ones (indexed PRNG), then only one
instance  of the PRNG would be needed to calculate the private keys for all the leaves. No PRNG's currently recommended by NIST~\cite{key17} have
this property. For both, the log and the fractal algorithms, solutions exist that use a PRNG which calculates the leaves' private keys in sequential
order (continuous PRNG). This requires storing multiple internal states of the continuous PRNG during the generation of the authentication paths. The
fractal algorithm stores as many continuous PRNG internal states as it has subtrees, whereas the log algorithm stores two continuous PRNG internal
states per $TreeHash$~\cite{key15} plus one for calculating the leaves that are left nodes. Our algorithm uses the same PRNG-approach as the fractal
one. When our algorithm skips a leaf calculation (because it would not contribute to the calculation of a right node stored in a subtree, see
Sec.~\ref{sec:overview}), it still calculates the leaf's private key and thus advances the state of the PRNG. Therefore, our algorithm and the
fractal one, store $L$ additional continuous PRNG states, whereas the log algorithm needs to store $2\times(H-K) + 1$ continuous PRNG
states~\cite{key15}. For the space analysis we choose the state size of the PRNG equal to the output size of the hash function used.

\subsection{The \textsl{TreeHash} Algorithm}\label{sec:tree_hash}
In this section we will explain the reason why we use the same $TreeHash$ scheduling as in~\cite{key15} together with the improved $TreeHash$
from~\cite{key15} and what impact this has on the performance of the algorithm. A $TreeHash$ instance which calculates a node on height $i$ and all
its children, is called $TreeHash_i$. For each $Desired$ tree in a subtree we need a lower $TreeHash_{bottomLevel}$ instance. Each of these instances
have up to $bottomLevel+1$ nodes on their stack. If we compute them simultaneously as it is done in~\cite{key13}, it can happen that each instance
has its maximum amount of nodes on their stack. The update scheduling algorithm from~\cite{key15} uses less space by computing the $TreeHash$
instances in a way, that at any given round the associated stacks are mostly empty~\cite{key8}. The basic idea is to start a freshly initialized
$TreeHash_{k}$ only if there is no $TreeHash$ with nodes of height smaller than $k$ on their stack. This is achieved by assigning each update to the
$TreeHash$ instance with the smallest tail height (see Algorithm~\ref{tailHeight}). The tail height is the smallest height for which there is a node
on the stack of the $TreeHash$. A terminated $TreeHash_k$  is considered to have an infinite tail height and an empty one is considered to have a
tail height of $k$. Furthermore, the improved $TreeHash$ from~\cite{key15} we use, changes the definition of a step as compared to the classical one.
A $step_C$ was originally considered in~\cite{key13} as either calculating a leaf node or an inner node. This is fine as long as a hash computation can
be considered to be as expensive as a leaf calculation. More often though, a leaf computation is significantly more expensive than the computation of
an inner node. This leads to a larger difference between the average and worst case time needed for a $step_C$. In~\cite{key15}, a $step_I$ consists of one
leaf's calculation and of as many inner node computations as possible before needing a new leaf, instead of processing just one leaf or one inner
node as in the classic case (see Algorithm~\ref{algo:treehash}).

\subsubsection*{Nodes' supply for the higher \textsl{TreeHash}}\label{sec:node_ready}
We wish to prove, that when we spend $L-1$ updates on the lower $TreeHash$ (see Alg.~\ref{treehash}), it produces nodes before the higher $TreeHash$
needs them for computing nodes in the $Desired$ tree. To prove this we use the same approach as in~\cite{key15}. We focus on a subtree $ST_k$ with a
lower $TreeHash_h$ (bottom level of $ST_k$ is $h$). We consider a time interval starting at the initialization of $TreeHash_h$ and ending at the time
when the next node at height $h$ is required by the higher $TreeHash$ of $ST_k$. We call this node $Need_h$. The higher $TreeHash$ is updated every
$2^h$ rounds and requires a bottom level node in each update. This means that in the time considered we execute $(L-1)\times 2^h$ updates. A higher
$TreeHash$ of a subtree on a lower level needs new nodes more frequently, because their authentication nodes change more often. For any given
$TreeHash_i$ with $i < h \:,\: \frac{2^h}{2^i}$ nodes are needed during the time interval defined above: $2^i$ updates are used up to complete a node
on height $i$. Therefore $TreeHash_i$ requires $\frac{2^h}{2^i} \times 2^i = 2^h$ updates to produce all needed nodes. If there are $N$ $TreeHash_i$
with $i < h$, then all of them together need at most $N\times 2^h$ updates to compute all their nodes. They may need less, because they may already
have nodes on their stack. There may be a partial contribution to any $TreeHash_j$ with $j
> h$. But they can only receive updates as long as they have nodes at height $< h$ (tail height $< h$). A $TreeHash_j$ needs at most $2^h$ updates to
raise its tail height to $h$. There are $L-N-2$ $TreeHash_j$ with $j > h$ (the top subtree has no $TreeHash$). Together they need at most
$(L-N-2)\times 2^h$ updates. All $TreeHash_{k}$ with  $k \neq h$ need at most $(L-N-2)\times 2^h + N\times 2^h = (L-2)\times 2^h$ updates. This
leaves $2^h$ updates for $TreeHash_h$, which are enough to compute $Need_h$.

\subsubsection*{Space and time analysis for the lower TreeHashes}
In~\cite{key15}, it was shown that when the improved scheduling algorithm is used with  $n \times \frac{1}{2}$ updates per round, a $TreeHash_l$
terminates at most $2^{l+1}$ rounds after its initialization ($n$ corresponds to the actual number of $TreeHash$ instances). This is clearly enough
for the log algorithm, because the authentication path needs a new right node on level $l$ every $2^{l+1}$ rounds. For our algorithm the higher
$TreeHash$ needs a new node every $2^{l}$ rounds which is twice as often. We thus need to distribute twice as many updates per lower $TreeHash$
instances with the improved scheduling algorithm from~\cite{key15}. That means $L-1$ updates per round in total.

\noindent In addition, when the improved scheduling algorithm is used to calculate nodes with a set of $TreeHash_i$ (where all $i$'s are different),
these nodes can share a stack~\cite{key15}. The amount of space needed by this shared stack is the same as that of the $TreeHash_i$ with the highest
$i$~\cite{key15}. Since the highest subtree (bottom level: $H-h$) does not have a lower $TreeHash$ instance, the highest level on which any node has
to be computed by a lower $TreeHash$ is the bottom level of the second highest subtree (with bottom level: $H-h-h$). So, the shared stack of our
algorithm stores at most $H-2h$ hash values.

\subsection{The space and time gains of our approach}~\label{sec:st_summ}
In this section we will give the total space and time bounds of our algorithm, and compare them with the log and fractal ones under the condition
that a continuous PRNG with an internal state equal in size of the hash value is used. We obtain the total space of our algorithm by summing up the
contributions of its different parts: $L\times(2^h-1)+H$  from the subtrees and authentication path (see Sec.~\ref{sec:pace_des}), $H-2h$ from the
lower $TreeHash$es (see Sec.~\ref{sec:tree_hash}) and $L$ from the PRNG internal states (see Sec.~\ref{sec:prng}). This sums up to $L\times
2^h+2H-2h$ times the hash value size.

\noindent For the time analysis we look at the number of leaves' calculations per round. The improved $TreeHash$ makes one leaf calculation per
update and we make at most $(L-1)$ lower $TreeHash$ updates per round. The higher $TreeHash$ never calculates leaves. So in the worst case all
$TreeHash$es together need $(L-1)$ leaves' calculations per round. We need an additional leaf calculation every two rounds to compute the left nodes
as shown in~\cite{key15}. Thus we need $L$ leaves' calculations per round in the worst case. In the average case however, we need less, as the first
node of the $2^h$ bottom level nodes of a $Desired$ tree is not computed, since it is not needed to compute any right node in the $Desired$ tree.

This reduces the average--case time by a factor $\frac{2^h-1}{2^h}$ and leads to a total of $\frac{2^h -1}{2^h} \times (L-1) + \frac{1}{2}$ leaves'
computations per round. The term $\frac{1}{2}$ enters the expression because the left node computation needs a leaf every two rounds. The average
case time bound holds true for only the first $2^H - 2^{H-h}$ rounds. Thereafter less leaf computations would be needed on average, because some
subtrees no longer need a $Desired$ tree. Table~\ref{tab:1} summarizes the above results and  Table~\ref{tab:2} does the same for the log space-- and
fractal--algorithm when a continuous PRNG with an internal state equal to the size of a hash value used.
\begin {table}[H]
    \begin{center}
        \begin{tabular}{ | l | c | c | c |}
            \hline
             \textsl{Bounds} & $h=1$, $L=H$ & $h=2$, $L=\frac{H}{2}$ & $h=log(H)$, $L=\frac{H}{log(H)}$\\ \hline
            Worst case: space  & $4H-2$ & $4H-4$ & $\frac{H^2}{log(H)} +2H -2log(H)$\\ \hline
            Average case: time & $\frac{H}{2}$ & $\frac{3H-2}{8}$ & $\frac{H-1}{log(H)} + \frac{1}{2}$ \\ \hline
            Worst case: time  & $H$ & $\frac{H}{2}$ & $\frac{H}{log(H)} - 1$\\ \hline
        \end{tabular}   \caption{Space--time trade--off of our Merkle tree traversal algorithm as a function  of $H$ (height of the tree) with $h$ (height of a subtree) as parameter.}\label{tab:1}
    \end{center}
\end {table}

\begin {table}[H]
    \begin{center}
        \begin{tabular}{ | l | c | c |}
            \hline
             \textsl{Bounds} & Log~\cite{key15}  $K = 2$ & Fractal~\cite{key3} $ h = \log(H)$ \\ \hline
            Worst case: space  & $5.5H-7$ & $\frac{5H^2 +2H}{2log(H)}$ \\ \hline
            Average case: time  & $\frac{H}{2} - \frac{1}{2}$ & $\frac{H}{log(H)} - 1$\\ \hline
            Worst case: time  & $\frac{H}{2}$ &  $\frac{2H}{log(H)} - 2$\\ \hline
        \end{tabular}
        \caption{Space--time trade--off of log algorithm~\cite{key15} and fractal algorithm~\cite{key3} optimized for storage space.
                 The values in the Table include the space needed by the continuous PRNG.}\label{tab:2}
    \end{center}
\end {table}

When $h=2$ our algorithm has better space and time bounds (or at least as good as in the case of worst--case time) than the log
algorithm~\cite{key15}. When we choose the same space--time trade-off parameter as in the fractal algorithm~\cite{key3} (column $h = log(H)$ in
Table~\ref{tab:1}), our algorithm needs less storage space.

\section{Implementation}
There are several aspects which are by purpose unspecified by the Merkle tree traversal algorithms. These are the hash function, the deterministic
pseudo--random number generator and the algorithm used for the leaf calculation. The latter is defined by the usage of the tree. Although the hash
function and PRNG are independent of the trees' usage, both have an impact on the cryptographic strength and the performance. The hash function used
for the traversal algorithm must be collision-resistant as shown in~\cite{key21}.   Thus the main selection criteria for the hash function are good
performance and strong security. A suitable candidate is BLAKE~\cite{key7}.

As a PRNG we chose an algorithm based on a hash function. This choice has the advantage that we do not need another cryptographic primitive.
In~\cite{key17}, NIST has recommended two continuous hash based PRNG's named HASH\_DBRG and HMAC\_DBRG.  Both of them have an internal state composed
of two values with the same length as the output length of the used hash function. HASH\_DBRG has the advantage that one of its two internal values
solely depends on the seed and does not change until a reseeding occurs. For Merkle trees, there is no reseeding necessary as long as less than
$2^{48}$ leaves exist~\cite{key17}. Hence, in our application one of its two internal values is the same for all used HASH\_DBRG instances within the
same Merkle tree. We prefer  HASH\_DBRG over HMAC\_DBRG because it uses less space and is more performant.

\section{Results}\label{sec:res}
We compared the performance of our algorithm with both, the log algorithm from~\cite{key15} and the fractal algorithm from~\cite{key3}. We chose as
performance parameters the number of leaf computations and the number of stored hash values. This choice is reasonable because the former is the most
expensive operation if the Merkle tree is used for signing, and the latter is a good indicator of the storage space needed. Operations like computing
a non--leaf node or generating a pseudo--random value have nearly no impact on the performance in the range of $H$ values of practical interest. A
leaf computation is exactly the same in each of the three algorithms and therefore only dependent on the underlying hardware for its performance.

To be able to present cogently the results, each data point represents an aggregation over $2^{10}$ rounds. Recall that one round corresponds to the
calculation of one authentication path. In the case of storage measurements one point represents the maximal amount of stored hash values at any time
during these $2^{10}$ rounds. In the case of the leaf computation one point represents the average number of leaves' computations done in one round
during the $2^{10}$ rounds.

\begin{figure}[h]
\includegraphics[scale=0.4, keepaspectratio=true]{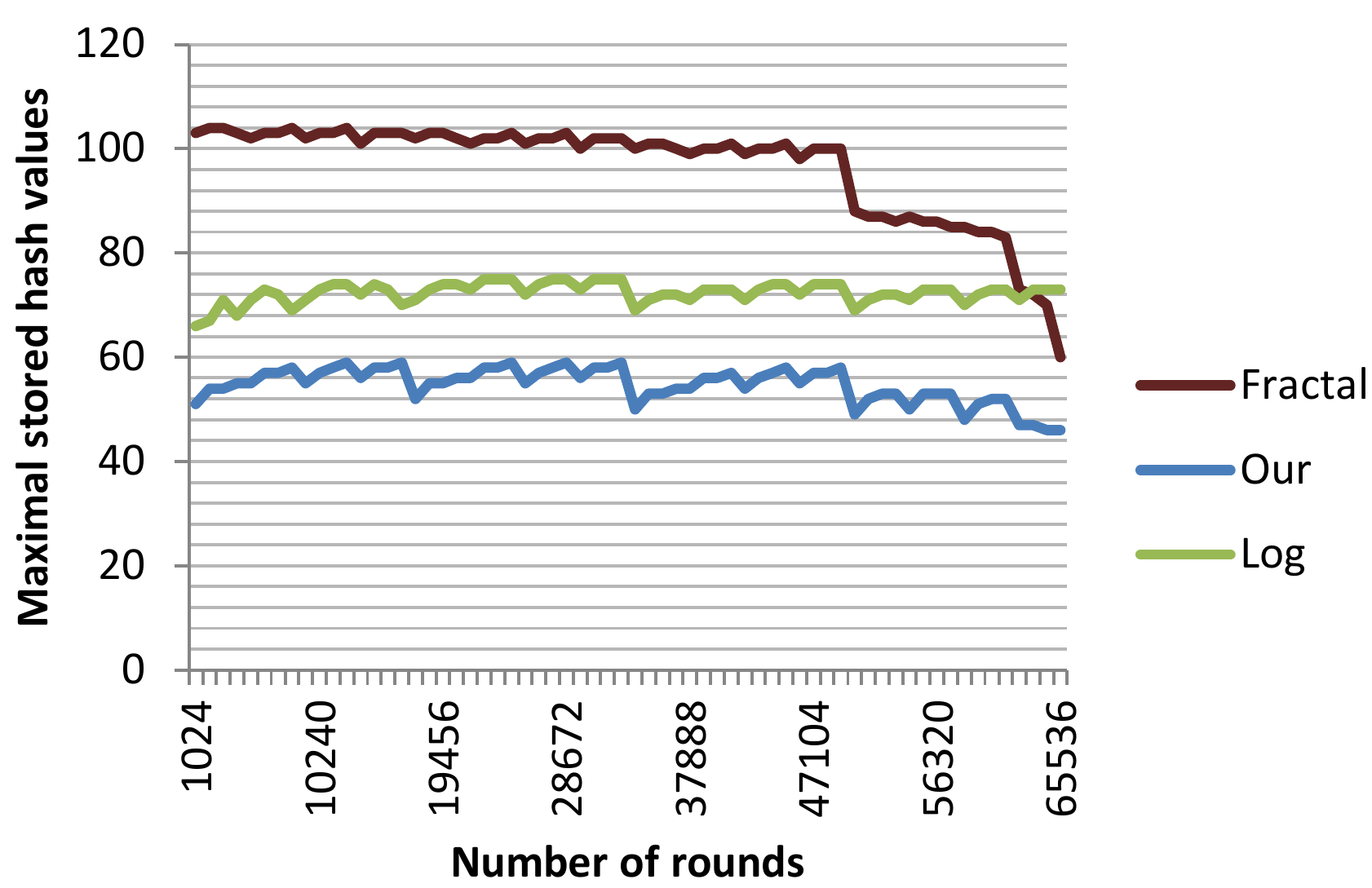}
\includegraphics[scale=0.4, keepaspectratio=true]{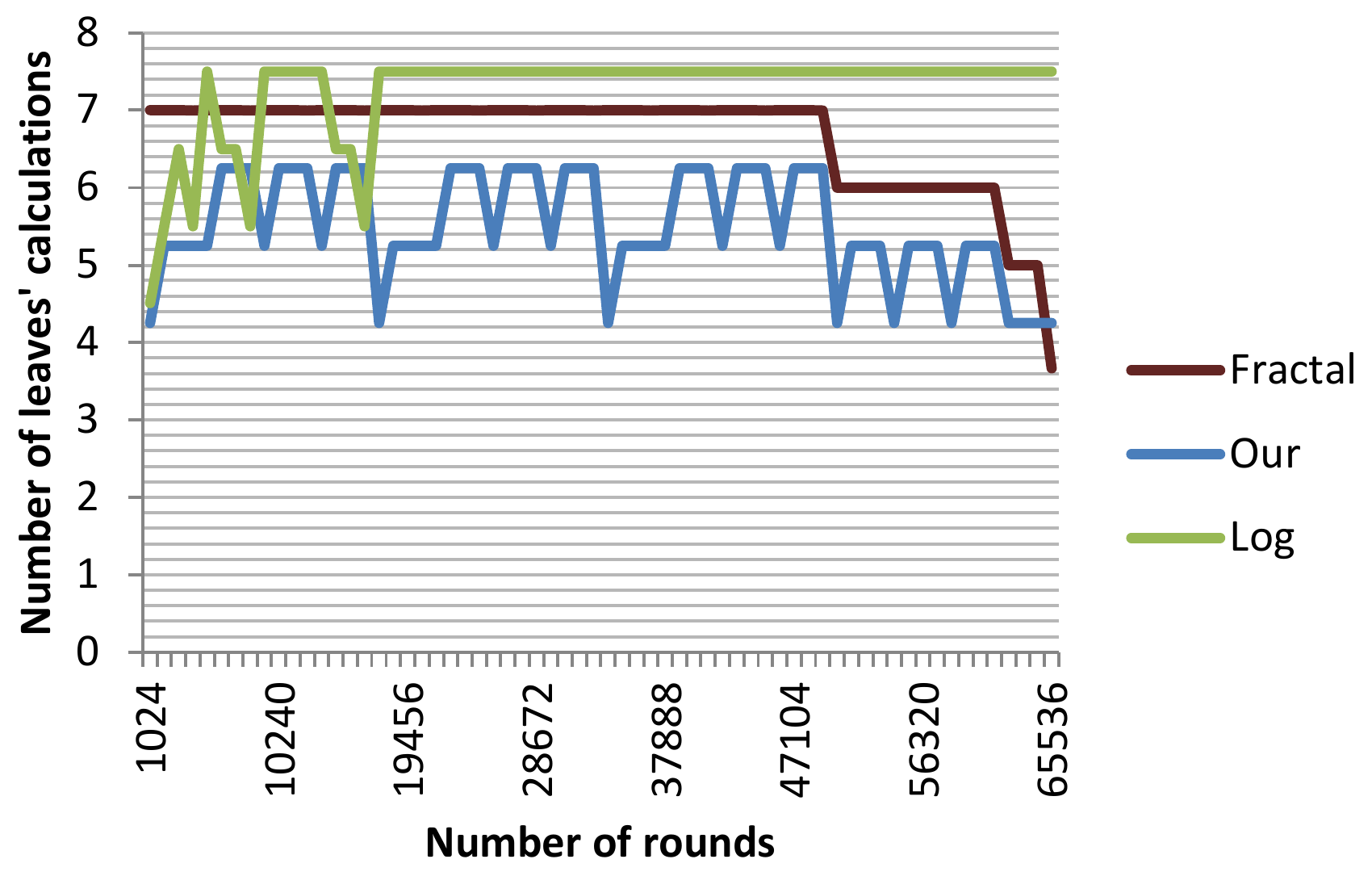}
\caption{\emph{Left}: Set one: Maximal number of stored hash values as a function of rounds for minimal space.
         \emph{Right}: Number of calculated leaves  as a function of rounds for minimal space. Parameters: $H=16$, $h=2$ and $K=2$. HASH\_DBRG is used as
         pseudo--random number generator. One round corresponds to the calculation of one authentication path.}\
    \protect\label{fig:6}
\end{figure}

\begin{figure}[h]
    \includegraphics[scale=0.4, keepaspectratio=true]{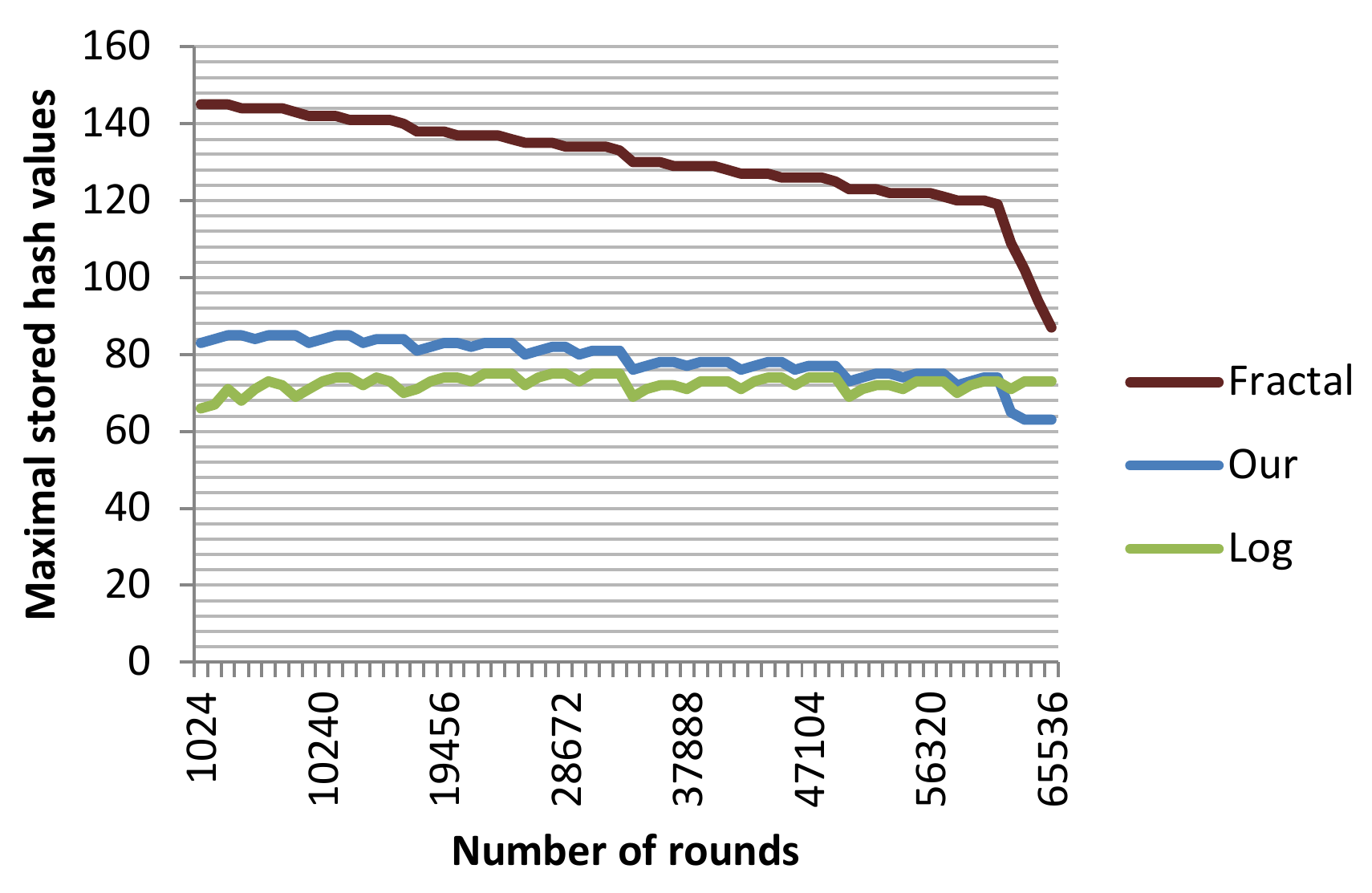}
    \includegraphics[scale=0.4, keepaspectratio=true]{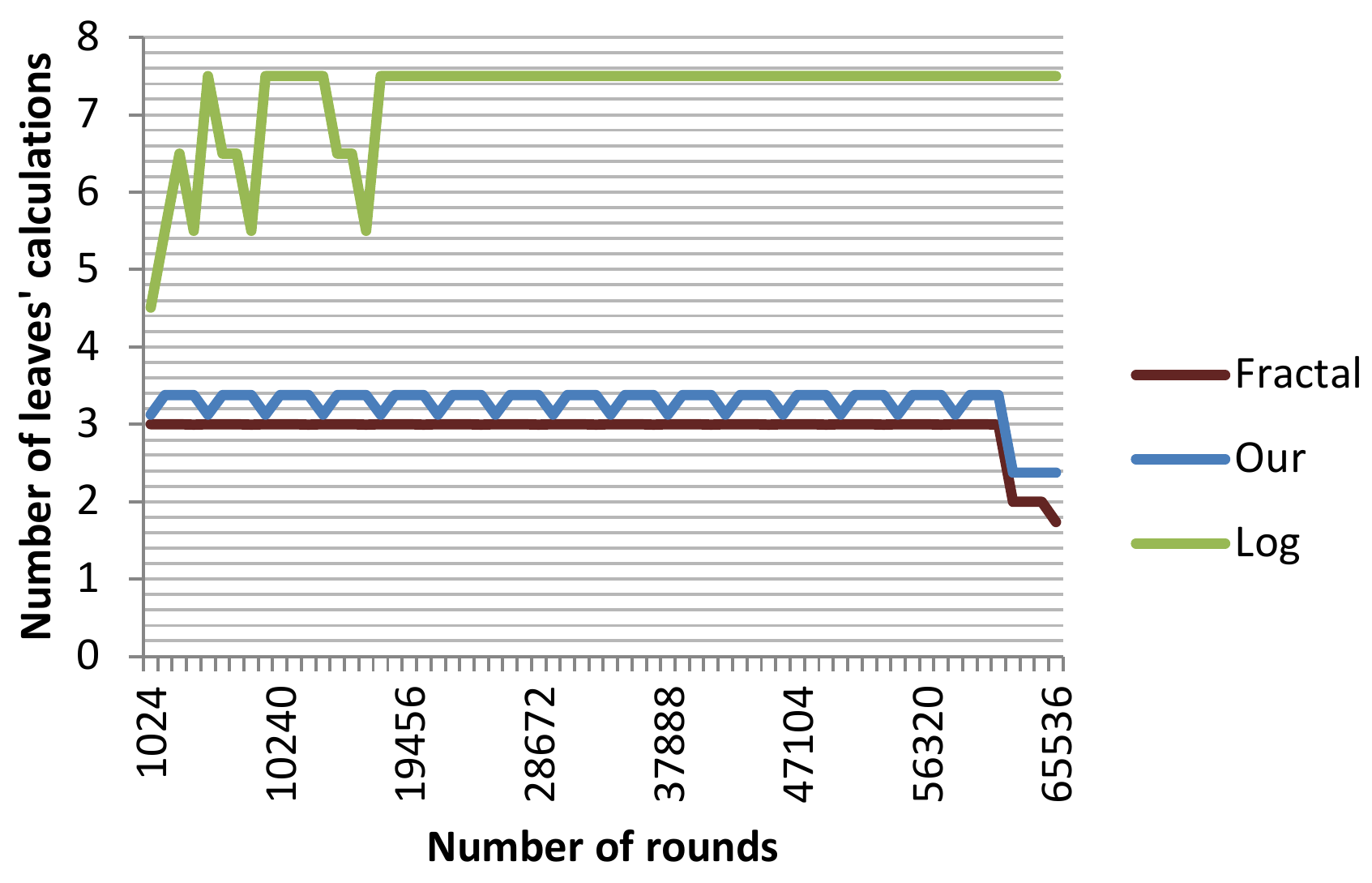}
    \caption{\emph{Left}: Set two: Number of stored hash values as a function of rounds  for similar space--time trade-off.
    \emph{Right}: Maximal number of calculated leaves as a function of rounds for similar space--time trade-off. Parameters: $H=16$, $h=4$ and $K = 2$.
            HASH\_DBRG is used as pseudo--random number generator. One round corresponds to the calculation of one authentication path.}\
    \protect\label{fig:5}
\end{figure}

\noindent  We will present the results for two sets of measurements with $2^{16}$ leaves. For the first set we choose the parameter such that each
algorithm uses its minimal space. For our and the fractal algorithm the minimal space for $H=16$ is achieved with $h=2$. In the case of the log
algorithm we have set $K$ (defined in~\cite{key15}) to $2$ in order to achieve minimal space usage.   The second set uses $h=log(H)$ as it was
proposed in~\cite{key3}. For the fractal and our algorithm this means $h=4$ for $H=16$ and $K = 2$ for the log algorithm. The NIST recommendation
HASH\_DBRG is used as PRNG for both sets of measurements. The results of these measurements are shown in Fig.~\ref{fig:5} for a similar space--time
trade-off as the fractal tree and in Fig.~\ref{fig:6} for minimal storage space.

We see that in a setting where a good space--time trade-off is needed, our algorithm uses less space and slightly more leaf calculations than the
fractal algorithm (at most $\frac{1}{2}$ more on average per round). If a minimal space solution is needed, our algorithm with $h=2$ uses less space
and less leaf calculations than both the log and the fractal algorithm.

\begin{figure}[h]
\centering
\includegraphics[scale=0.7, keepaspectratio=true]{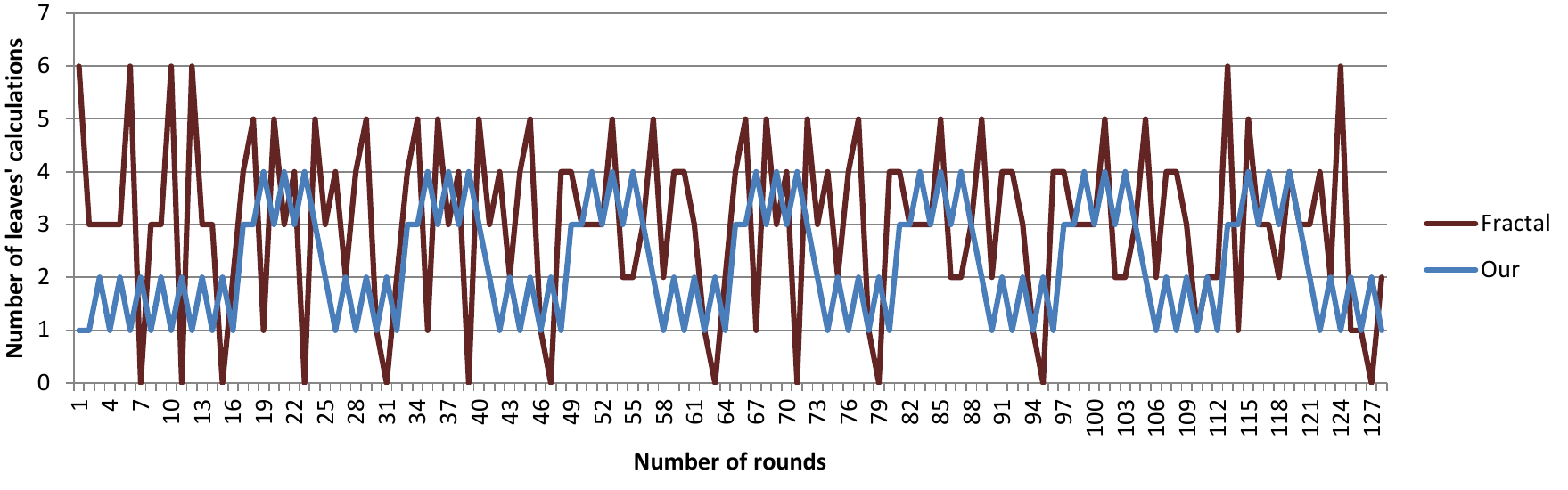}
\caption{Number of calculated leaves as function of rounds for similar space--time trade-off (first 128 rounds in detail).
         Parameters: $H=16$ and $h=4$. HASH\_DBRG is used as pseudo--random number generator. One round corresponds to the calculation of one authentication path.}\
    \protect\label{fig:7}
\end{figure}

\noindent In addition, the plots show  a weak point of our algorithm compared with the log algorithm: the number of leaves' calculations is not
constant. The fractal algorithm for similar parameter shows even greater fluctuations, but they are not visible in Fig.~\ref{fig:5}, because they
cancel each other out over the $2^{10}$ rounds. If we measure the first $2^{7}$ rounds with no aggregation we see that the deviations of our
algorithm decrease markedly (see Fig.~\ref{fig:7}) compared to the fractal one.

\noindent The full package with source code and measurements results is available at~\cite{key0}.

\section{Conclusion}

We developed an algorithm for the Merkle tree traversal problem which combines the efficient space-time trade-off from the fractal algorithm and the
space efficiency from the log algorithm. An exhaustive analysis of the space and time efficiency of our algorithm in function of the parameters $H$
 and $h$  has shown that if a continuous PRNG
is used, our algorithm has a space advantage over the log and fractal algorithms and a time advantage over the log algorithm.

We further programmed a low storage--space and a low time--overhead version of the algorithm in Java and measured its performance with respect to the
two different implementations. Our implementation needs about a factor 2 less space than the fractal algorithm, when minimum space is required.

Ours as well as the log and fractal algorithms suffer from a long initialisation time for large values of $H$. This problem was solved by the
CMSS~\cite{key18} and GMSS~\cite{key19} algorithms. These two algorithms use a stacked series of Merkle trees where the higher trees sign the roots
of their child trees and the lowest tree is used for the real cryptographic purpose. Both of them thus rely on a solution of the Merkle traversal
problem for each layer, for which our algorithm could be used instead of the current ones. It is possible to use different parameters for different
layers in the CMSS or GMSS. In addition, the higher trees used in these schemes favor Winternitz as leaf calculation function which is significantly
more expensive than an inner node computation, and thus can profit from the improved $TreeHash$ used in our algorithm. The XMSS~\cite{key20} is an
extension to the Merkle signature scheme (MSS) which allows to use a hash function which is only second-pre--image resistant instead of collision
resistant. It is based on the log algorithm and the usage of a forward secure continuous PRNG. Under these circumstances, our algorithm would be a
good replacement for the log algorithm: it would use less space and provide greater flexibility.

\section*{Acknowledgements}
This work was partially funded by the Hasler Foundation Grant no. 12002-- "An optimized CPU architecture for cryptological functions". We thank an
anonymous reviewer for insightful remarks.

\appendix
\section{Appendix}\label{app}
\subsection{Algorithms}
The algorithm descriptions use an oracle for the leaves' computations. The oracle gets the leaf's index as input. We should modify the algorithms (as
explained in Sec.~\ref{sec:prng}) in the case that the leaf's computation is based on a continuous PRNG and it needs a private key as input.

Our algorithm uses the following data structures:
\begin{enumerate}
    \item $Auth_h$, $h = 0, . . . ,H-1$. An array of nodes that stores the current authentication path.
    \item $Subtree_h$, $h = 0, . . . ,L-1$. An array of subtree structures with the following properties:
    \begin{enumerate}
                \item bottomLevel: the minimal height for which the subtree stores nodes.
                \item rootLevel: the height of the root of the subtree.
                \item tree: the data structure for the $Exist$ and $Desired$ tree with the following functions:
                                        \begin{enumerate}
                                                \item get($j,k$): get $k$th node (from left to right) with height $j$ in the subtree
                                                \item add(node): store node in the subtree
                                                \item remove($j,k$): remove $k$th node (from left to right) with height $j$  in the subtree
                                        \end{enumerate}
                \item stackHigh: the stack for the higher $TreeHash$.
                \item nextIndex: the index of the next leaf needed by the lower $TreeHash$.
                \item bottomLevelNode: the node of lower $TreeHash$ which is stored outside the shared stack~\cite{key15}.
                \item stackLow: the stack for the lower $TreeHash$ (the part of the shared stack currently containing nodes for this
                $Subtree$~\cite{key15}).
        \end{enumerate}
        \item $Lea\!fCalc(i)$, $i = 0, . . . ,2^{H}-1$. Oracle for calculating the leaf $i$.
\end{enumerate}

Our algorithm has the following phases:
\begin{enumerate}
    \item Init: $TreeHash$ computes the root node. During this process it stores right nodes of the left--most $Exist$ trees and the nodes of the first authentication
    path(Algorithm~\ref{algo:init})
        \item Generation of the authentication paths: repeat $2^H$ times:
        \begin{enumerate}
                \item Output current authentication path $Auth_h$, $h = 0, . . . ,H-1$
                \item Update lower $TreeHash$es (Algorithm~\ref{algo:2})
                \item Compute next authentication path (Algorithm~\ref{algo:1})
        \end{enumerate}
\end{enumerate}

\begin{algorithm}[h]
\caption[Key generation]{Key generation (PK) and Merkle tree setup.}
  \begin{framed}
  \begin{algorithmic}
  \STATE{$ \mathbf{INPUT}: $}
  \STATE $ \mathbf{OUTPUT}: \mathrm{PK}$

  \COMMENT{Initialize L-1 subtrees}
  \FORALL{ $\mathrm{Subtree_i}$ with $i \in \{0,\cdots,L-1\}$}
   \STATE{$\mathrm{Subtree_i.tree} \leftarrow empty$}
   \IF{i < (L-1)}
     \STATE{$\mathrm{Subtree_i.stackHigh} \leftarrow empty$}
     \STATE{$\mathrm{Subtree_i.stackLow} \leftarrow empty$}
   \ENDIF
   \STATE{$\mathrm{Subtree_i.bottomLevel} \leftarrow i \times h$}
   \STATE{$\mathrm{Subtree_i.rootLevel} \leftarrow \mathrm{Subtree_i.bottomLevel}+ h$}
   \STATE{$\mathrm{Subtree_i.nextIndex} \leftarrow 2^{Subtree_i.rootLevel} - 1$}
  \ENDFOR

  \COMMENT{Initialize Stack, set $\mathrm{Leaf\:level}\: k = 0$}
  \STATE $k \leftarrow 0$
    \STATE $Stack \leftarrow empty$
    \STATE $Stack.push(LeafCalc(k))$
    \STATE $k \leftarrow k+1$
  \WHILE{$\mathrm{Stack.peek.height} < H$}
    \STATE $\mathrm{TreeHash(Stack, LeafCalc(k), Process_0,null)}$
    \STATE $k \leftarrow k+1$
  \ENDWHILE
  \STATE $PK \leftarrow \mathrm{Stack.pop}$
  \STATE \textbf{return}  $PK$

\end{algorithmic}
\end{framed}
\label{algo:init}
\end{algorithm}

\begin{algorithm}[h]
\caption{$\mathrm{TailHeight}$: Calculation of the height of the lowest node on a $\mathrm{stackLow}$}
    \begin{framed}
  \begin{algorithmic}
  \STATE $ \mathbf{INPUT}: \: \mathrm{subtree \: index} \: i$
  \STATE $ \mathbf{OUTPUT}: height$
    \IF{$\mathrm{Subtree_i\: has\: a\: stackHigh} \wedge \:\lnot(\mathrm{Subtree_i.bottomLevelNode})$}
        \IF{$\mathrm{Subtree_i.stackLow} == \mathrm{empty}$}
            \STATE $height \leftarrow  \mathrm{Subtree_i.bottomLevel}$
        \ELSE
            \STATE $height \leftarrow  \mathrm{Subtree_i.stackLow.tosNode.height}$
        \ENDIF
    \ELSE
        \STATE $height \leftarrow  \infty$
    \ENDIF

    \STATE \textbf{return}  $height$
\end{algorithmic}
\end{framed}
\label{tailHeight}
\end{algorithm}

\begin{algorithm}[h]
\caption{$\mathrm{Process_0}$}
    \begin{framed}
  \begin{algorithmic}
  \STATE $ \mathbf{INPUT}: \mathrm{Node}, \mathrm{index}\: j$
  \STATE $ \mathbf{OUTPUT}: continue$
  \IF{$\mathrm{Node.index} \leq 2^{\mathrm{SubTreeForLevel(Node.height).rootLevel} - \mathrm{Node.height}} \wedge \mathrm{Node.index} \pmod 2 == 1$}
    \STATE $\mathrm{SubTreeForLevel(Node.height).tree.add(Node)}$
  \ENDIF
  \IF{$\mathrm{Node.index} == 1$}
    \STATE $\mathrm{Auth_{Node.height}} \leftarrow \mathrm{Node}$
  \ENDIF
  \STATE \textbf{return} $continue \leftarrow 1$
\end{algorithmic}
\end{framed}
\label{proc0}
\end{algorithm}

\begin{algorithm}[h]
\caption{$\mathrm{Process_1}$: }
    \begin{framed}
  \begin{algorithmic}
  \STATE $ \mathbf{INPUT}: \mathrm{Node}; \: \mathrm{subtree \: index} \: j$
  \STATE $ \mathbf{OUTPUT}: continue$

  \IF{$\mathrm{Node.height} == Subtree_j.bottomLevel$}
    \STATE $\mathrm{Subtree_j.bottomLevelNode} \leftarrow \mathrm{Node}$
    \STATE $continue \leftarrow  0$
  \ELSE
    \STATE $continue \leftarrow  1 $
  \ENDIF
    \STATE \textbf{return}  $continue$
\end{algorithmic}
\end{framed}
\end{algorithm}

\begin{algorithm}[h]
\caption{$\mathrm{Process_2}$:}
\begin{framed}
\begin{algorithmic}
  \STATE $ \mathbf{INPUT}: \mathrm{Node};\: \mathrm{subtree \: index} \: i$
  \STATE $ \mathbf{OUTPUT}: continue$
  \IF{$\mathrm{Node} \neq dummy $}
        \STATE $continue \leftarrow  1 $
    \IF{$\mathrm{Node.index} \pmod 2 == 1 $}
      \STATE $\mathrm{Subtree_i.tree.add(Node)}$
         \IF {$\mathrm{Node.index}/2 \pmod{2^{\mathrm{Subtree_i.rootLevel} - \mathrm{Node.height} -1}} == 0$}
           \STATE $continue \leftarrow  0$
         \ENDIF
    \ENDIF
         \IF{$Node.height == \mathrm{Subtree_i.rootLevel} -1$}
           \COMMENT{Current $Desired$ tree becomes new $Exist$ tree}
                     \IF{$\mathrm{Subtree_i.nextIndex}+1 >= 2^H$}
                             \COMMENT{It was the last $Desired$}
                             \STATE{ $\mathrm{Subtree_i.stackHigh} \leftarrow remove$}
                     \ENDIF
         \ENDIF
  \ELSE
    \STATE $continue \leftarrow  0$
  \ENDIF
    \STATE \textbf{return}  $continue$
\end{algorithmic}
\end{framed}
\label{proc2}
\end{algorithm}

\begin{algorithm}[h]
\caption{Distribution of  updates to the active lower $TreeHash$ instances:}
  \begin{framed}
  \begin{algorithmic}
  \STATE $ \mathbf{INPUT}:\: \mathrm{leaf}\: \mathrm{index}\: i \in \{1, \cdots, 2^H -1\}$

  \STATE $updates  \leftarrow $  number of desiredTree in SubTrees
  \REPEAT

        \COMMENT{Find TreeHash instance with lowest tail height, on a tie use the one with lowest index}
    \STATE $s \leftarrow min\{ l: \forall \: \mathrm{TailHeight(l)} == \underset{j = 0, \cdots, L-2}{min}\{\mathrm{TailHeight(j)}\}\}$
    \STATE $\mathrm{Subtree_s.nextIndex} \leftarrow \mathrm{Subtree_s.nextIndex}+1$
    \IF{$\mathrm{Subtree_s.nextIndex} \pmod {2^{\mathrm{Subtree_s.rootLevel}}} \geq 2^{\mathrm{Subtree_s.bottomLevel}}\:\:\:\:$}
      \STATE $TreeHash(\mathrm{Subtree_s.stackLow}, LeafCalc(\mathrm{Subtree_s.nextIndex}), Process_1, s)$
    \ELSE
      \IF{$\mathrm{Subtree_s.nextIndex}+1 \pmod {2^{\mathrm{Subtree_s.rootLevel}}} == 2^{\mathrm{Subtree_s.bottomLevel}}$}
                \STATE {$\mathrm{Subtree_s.bottomLevelNode} \leftarrow dummy$}
            \ENDIF
    \ENDIF
    \STATE $updates \leftarrow updates-1$
  \UNTIL $updates == 0$

    \end{algorithmic}
    \end{framed}
    \label{algo:2}
 \end{algorithm}

\begin{algorithm}[h]
    \caption{Generation of the next authentication path. ($\mathrm{SubTreeForLevel(l)}$ is the Subtree containing level l.)}
    \begin{framed}
    \begin{algorithmic}
    \STATE $ \mathbf{INPUT}:\: \mathrm{leaf}\: \mathrm{index}\: i \in \{1, \cdots, 2^H -1\}$

  \COMMENT{$k$ is 0 if leaf $i$ is a righ node and $k\neq 0$ means the height of the first parent of leaf $i$ that is a right node}
  \STATE $k \leftarrow \underset{m = 0, \cdots, H}{max}\{m: i \mod 2^m == 0 \} $
  \IF{$k == 0$}
    \STATE {$\mathrm{Auth_0 \leftarrow \mathrm{LeafCalc(i-1)}}$}
  \ELSE
        \STATE $\mathrm{leftNode} \leftarrow \mathrm{Auth_{k-1}}$
        \STATE {$\mathrm{rightNode} \leftarrow \mathrm{SubTreeForLevel(k-1).tree.get(leftNode.index \oplus 1,k-1)}$}
    \STATE {$\mathrm{Auth_{k}} \leftarrow \mathrm{hash(leftNode || rightNode)}$}
    \STATE {$\mathrm{SubTreeForLevel(k-1).tree.remove(j,k-1)}$}
  \ENDIF

    \COMMENT{Remove sibling of $\mathrm{Auth_{k}}$}
  \IF {$\mathrm{Auth_{k}.index/2} \pmod 2 == 1$}
    \STATE {$\mathrm{SubTreeForLevel(k).remove(Auth_{k}.index \oplus 1,k)}$}
  \ENDIF

    \COMMENT{Run through stackHigh in all Subtrees whose $Auth_{bottomLevel}$ changed}
  \FORALL {$r \in \{0\cdots L-2\}$ where $\mathrm{Subtree_r.bottomLevel} \leq k$}
    \IF{SubTree[r] has a desiredTree}
      \STATE $TreeHash(\mathrm{Subtree_r.stackHigh}, \mathrm{Subtree_r.bottomLevelNode}, Process_2, r)$
      \STATE $\mathrm{Subtree_r.bottomLevelNode} \leftarrow remove $
    \ENDIF
  \ENDFOR
  \FORALL {$t \in \{0\cdots k-1\}$ }
    \STATE $\mathrm{Auth_{t}}  \leftarrow \mathrm{SubTreeForLevel(t).tree.get(n,(i/2^t) \oplus 1)}$
  \ENDFOR
    \STATE \textbf{return} $\mathrm{Auth_{j}}\: \forall j \in \{0, \cdots, H-1\}$
\end{algorithmic}
\end{framed}
\label{algo:1}
\end{algorithm}

\newpage

\clearpage

%
%

\clearpage
\end{document}